\documentclass[conference]{IEEEtran}
\IEEEoverridecommandlockouts
\usepackage{cite}
\usepackage{amsmath,amssymb,amsfonts}
\usepackage{algorithmic}
\usepackage{graphicx}
\usepackage{textcomp}
\usepackage{xcolor}
\usepackage{marvosym}

\usepackage{booktabs}
\usepackage{multirow}
\usepackage{CJKutf8}
\usepackage{array}
\usepackage{stfloats}
\usepackage{tikz}
\usepackage{flushend}
\usepackage{lipsum}
\usepackage{listings}

\usepackage[most]{tcolorbox}
\tcbuselibrary{breakable} % 引入 breakable 库
\tcbuselibrary{skins} % 引入 skins 库

\definecolor{mygreen}{rgb}{0,0.6,0}
\definecolor{mygray}{rgb}{0.5,.0.5,0.5}
\definecolor{mymauve}{rgb}{0.58,0,0.82}

\def\BibTeX{{\rm B\kern-.05em{\sc i\kern-.025em b}\kern-.08em
    T\kern-.1667em\lower.7ex\hbox{E}\kern-.125emX}}
\begin{document}

\title{ARCEAK: An Automated Rule Checking Framework Enhanced with Architectural Knowledge}
\author{\IEEEauthorblockN{anonymous authors}
\IEEEauthorblockA{\textit{Submission ID: 48}}}
 \author{
     \IEEEauthorblockN{Junyong Chen\IEEEauthorrefmark{1},
     Ling-I Wu\IEEEauthorrefmark{1},
     Minyu Chen\IEEEauthorrefmark{1},
     Xiaoying Qian\IEEEauthorrefmark{2},
     Haoze Zhu\IEEEauthorrefmark{3},
     Qiongfang Zhang\IEEEauthorrefmark{3},
     Guoqiang Li\IEEEauthorrefmark{1}\,\Letter\thanks{\Letter\, Corresponding author}}
 \IEEEauthorblockA{
 \IEEEauthorrefmark{1}\textit{Shanghai Jiao Tong University, Shanghai 200240, China}\\
 \{chen.jy, edithwuly, minkow, li.g\}@sjtu.edu.cn\\
 \IEEEauthorrefmark{2}{\textit{Beijing University of Civil Engineering and Architecture, Beijing, China}}\\
 202304050101@stu.bucea.edu.cn\\
 \IEEEauthorrefmark{3}\textit{East China Architectural Design \& Research Institute, Shanghai, China}\\
 \{haoze\_zhu, qiongfang\_zhang\}@ecadi.com
  }
 }

\maketitle

\begin{abstract}
Automated Rule Checking (ARC) plays a crucial role in advancing the construction industry by addressing the laborious, inconsistent, and error-prone nature of traditional model review conducted by industry professionals. Manual assessment against intricate sets of rules often leads to significant project delays and expenses. In response to these challenges, ARC offers a promising solution to improve efficiency and compliance in design within the construction sector. However, the main challenge of ARC lies in translating regulatory text into a format suitable for computer processing. Current methods for rule interpretation require extensive manual labor, thereby limiting their practicality. To address this issue, our study introduces a novel approach that decomposes ARC into two distinct tasks: rule information extraction and verification code generation. Leveraging generative pre-trained transformers, our method aims to streamline the interpretation of regulatory texts and simplify the process of generating model compliance checking code. Through empirical evaluation and case studies, we showcase the effectiveness and potential of our approach in automating code compliance checking, enhancing the efficiency and reliability of construction projects.
\end{abstract}

\begin{IEEEkeywords}
	
Automated rule checking,
Rule interpretation,
Information extraction,
Code generation,
Generative pre-trained transformer,
Large language model
\end{IEEEkeywords}

\section{Introduction}

The Architecture, Engineering and Construction  (AEC)
industry has undergone significant digital transformation in recent years, transitioning from traditional 2D line drawings to data-centric construction processes. Ensuring compliance with established rules and regulations is crucial for delivering high-quality building design models. To replace or augment manual rule checking, the concept of \emph{Automated Rule Checking (ARC)} has been introduced as a potential solution. ARC refers to a technology-driven approach  automated building rule compliance checking by converting rules into machine-readable formats~\cite{ilal2017computer}.
% Research on ARC has been ongoing for nearly 70 years and has emerged as a prominent theme in construction-related studies~\cite{lin2006building, ma2015algorithm, xuejiao2019automatic}. However, existing ARC frameworks typically convert natural language building rules into structured formats, such as knowledge graphs, using fine-tuned language models like BERT~\cite{devlin2019bert} (\ccr{GQ: logically incorrect. BERT does not exist nearly 70 years}). Despite this approach, a large amount of domain-specific labeled data is still required to fine-tune a pre-trained language model.
Traditional ARC approaches often rely on hard-coded or manual rule interpretation methods~\cite{eastman2009automatic, jiang2015automated}, while modern researches benefit from advancements in natural language processing~\cite{lin2006building, ma2015algorithm, xuejiao2019automatic}. Recent ARC approaches leverage fine-tuned language models such as BERT~\cite{devlin2019bert} to convert natural language building rules into structured formats~\cite{zhou2022arccfg, zheng2022knowledge}, including knowledge graphs. Despite these advancements, the substantial demand for domain-specific labeled data for fine-tuning continues to hinder the large-scale deployment of language models in the ARC domain.

Very recently, large language models (LLMs) such as GPT-3.5-Turbo have shown remarkable performance in various natural language processing (NLP) downstream tasks, including sentiment analysis~\cite{wang2024chatgpt}, natural language inference~\cite{li2023chatgpt} and misinformation detection~\cite{parikh2023exploring}.
% Leveraging their massive parameter size and extensive textual training data, LLMs can adapt to domain specific tasks with simple task-specific prompt instructions, such as mental health analysis~\cite{lamichhane2023evaluation}.
% Some researchers~\cite{he2024llmbased, bouzenia2024repairagent} have also assigned different characters to specialize in various tasks and work as a team to develop software engineering systems.
Researchers~\cite{he2024llmbased, bouzenia2024repairagent} have even assigned different characters to these models to specialize in various tasks and collaborate on developing software engineering systems.
% (\ccr{bridge paragraph, finish this})
One of the most notable capabilities of LLMs is their zero-shot generalization across diverse tasks~\cite{DBLP:conf/nips/KojimaGRMI22}. With well-designed prompts, these models can perform complex tasks at a human-level ability, addressing issues related to the lack of training data for fine-tuning.
% The most impressive capability of LLMs is the zero-shot generalization among various tasks~\cite{DBLP:conf/nips/KojimaGRMI22}.
% % (cite llms is zero-shot learner paper here!!!!)
% Language models can perform human-level ability in complex tasks, provided with only well-designed prompts, which resolve the issues of lacking training data for fine-tuning language models.
As for the ARC task, LLMs have already demonstrates its effectiveness in similar tasks such as natural language understanding~\cite{DBLP:conf/iclr/WangSMHLB19, DBLP:conf/nips/WangPNSMHLB19} and code generation~\cite{thakur2023verigen, jiang2023selfplanning}. Thus, we also expect LLMs to address several challenges such as building rule interpretation and checking code generation.

To harness the powerful in-context learning capabilities of LLMs for the ARC task, we propose ARCEAK, an \textbf{A}utomated \textbf{R}ule \textbf{C}hecking framework \textbf{E}nhanced with \textbf{A}rchitectural \textbf{K}nowledge. Our framework consists of two main stages. The first stage is LLM-based Rule Information Extraction, which focuses on extracting verification-related information from building rules. This stage is further divided into two steps: entity discovery (ED) and event extraction (EE). ED involves recognizing construction domain-specific entities, while EE identifies assignments related to these entities using construction domain knowledge augmented prompt engineering.
The second one is LLM-based verification code generation, aiming to generate fine-grained, executable verification code by combining the extracted entities, events, and rule entry content with code generation prompts.
This stage is also divided into two steps: code framework generation and rule checking code completion. Code framework generation involves creating the skeleton of the verification code, and rule checking code completion fills in the specific execution code.
We conducted comprehensive experiments to evaluate the performance of our proposed framework. The results show that, for the Rule Information Extraction stage, ARCEAK has improved the F1 score of ED by 60\% and increased the precision of EE by 2.2\%. For the Verification Code Generation stage, ARCEAK has achieved a compile pass rate of 63\% with GPT-3.5-Turbo and a logic pass rate of 24\% with GPT-4-Turbo. Meanwhile, the knowledge-augmented code generation has demonstrated significantly better performance than non-knowledge-augmented code generation.

The contributions of this paper are summarized as followed:
\begin{itemize}
    \item We propose a novel LLM-based ARC framework enhanced with architectural knowledge, achieving an almost fully automated process for converting natural language building rules into executable verification code.
    \item We develop a robust construction domain-specific entity and event schema and established appropriate verification code generation granularity to balance problem complexity with the LLMs' code generation capabilities, ensuring the framework's scalability and transferability.
    \item We implement the ARCEAK framework and evaluated its performance using metrics designed with architecture experts, addressing both software engineering requirements and construction reliability.
\end{itemize}

\emph{Paper Organization:} The structure of the remaining part of this paper is as follows: In Section \ref{sec: preliminaries}, we offer an extensive review of the background and related research. The design and implementation of the entire ARCEAK process are detailed in Section \ref{sec: methodology}. Section \ref{sec: evaluation} presents the performance of different stages and the final results achieved by ARCEAK. Section \ref{sec: discussion} presents multiple case studies and analyzes potential threats to validity. Finally, conclusions are outlined in Section \ref{sec: conclusion}.
 % Finally, related work and conclusions are outlined in Sections \ref{sec: related work} and \ref{sec: conclusion}, respectively.

\section{Background}
% \section{Background and Related Work}

\label{sec: preliminaries}
In this section, we first introduce the background knowledge
related to LLM and prompt engineering. Then, we explore key concepts in information extraction. Finally, we discuss code generation techniques, including both specialized and general LLMs for code generation.
% In this section, we first introduce the background of large language models , including their training stages and prompt engineering techniques. Next, we cover information extraction methods using LLMs, followed by an overview of recent advancements in code generation leveraging LLMs.

% \subsection{Problem Definition}
% A building rule (also known as building code) consists of regulatory  text, tables, and formulas, which is semi-structured. Rule entities and events are obtained from textual rule sequences by extracting information using Large Language Models(LLMs).

% A set $R$ denotes the raw building rule, which includes $N$ regulatory sections.The $ith$ regulation section is the $ith$ list of rule contents $C_r$. The list of $C_r$ is separated and converted into list of textual rule sequences $S_r$. The objective of information extraction is to extract the target information $t_i$ (e.g.,entities,relations or events) from $S_r$.

% \subsection{Preliminaries}

%\subsection{Large Language Model}
\subsection{Large Language Model}
\textbf{Large Language Models (LLMs)} are generative models based on the pre-trained Transformer architecture\cite{vaswani2017attention}.
% Most LLMs utilize a generative model architecture, where given a sentence of $n$ tokens, the model is trained to maximize the likelihood of the ground-truth token $t_i$ at the current time step $t$ based on its preceding sequence $t_{i-1},...,t_1$.
% The training of LLMs typically follows three main processes: unsupervised training on large amounts of unlabeled text data without explicit human annotations, supervised fine-tuning on labeled data relevant to specific tasks or domains, and reinforcement learning on feedback from human annotators or evaluators.
Leveraging extensive multimodal data and employing pre-training and fine-tuning techniques, LLMs has significantly advanced the field of NLP by enhancing capabilities in multilingual translation~\cite{DBLP:conf/emnlp/ChronopoulouSF20, DBLP:conf/eacl/SticklandLG21, li2024eliciting}, summarization generation~\cite{zhang2024benchmarking, ravaut2024context}, and code simulation~\cite{chen2024language, lamalfa2024code}.
LLM training generally involves three key stages: unsupervised learning on vast amounts of unlabeled text data without direct human annotations, supervised fine-tuning using labeled data tailored to specific tasks or domains, and reinforcement learning based on feedback from human evaluators or annotators.

% \subsection{Prompt Engineering}

\textbf{Prompt Engineering} offers a powerful method to extend capabilities of LLMs without the need for extensive model retraining or modification of parameters and has emerged as a transformative technique in the realm of LLMs~\cite{radford2019language, wei2022chain}. This approach harnesses prompts to tailor model behavior to specific tasks or domains, thereby enhancing model efficacy and versatility. Prompt engineering can be categorized into two main types: \emph{zero-shot prompting}~\cite{brown2020language} and \emph{few-shot prompting}~\cite{DBLP:conf/nips/KojimaGRMI22}. The key distinction between these two lies in whether examples related to the task are provided to the language model. Zero-shot prompting involves giving the LLM an instruction without any examples, while few-shot prompting includes providing a few relevant examples to guide the LLM's response.
Zero-shot prompting and few-shot prompting can also be combined with in-context learning methods to further enhance the performance of LLMs in solving complex problems or unseen domain-specific tasks. One of the most widely-used in-context learning methods is Chain-of-Thought Prompting~\cite{wei2022chain}, which decomposes problems into intermediate steps and solves each one before arriving at the final answer.
\subsection{Information Extraction}
\textbf{Information Extraction (IE)} is a fundamental domain in NLP  to convert plain text into structured knowledge format~\cite{wang2023instructuie, lu2022unified}.The IE tasks cover:
% The IE tasks cover: \emph{entity discovery (ED)}~\cite{yan2021unified} and \emph{event extraction (EE)}~\cite{qi2023mastering}. ED encompasses both entity recognition~\cite{cui2021template} and entity typing~\cite{yuan2022generative}. The former is concerned with identifying spans of entities (e.g., ‘Steve’), and the latter focuses on assigning types to these identified entities (e.g., ‘PERSON’). EE involves event detection~\cite{veyseh2021unleash} and event argument extraction~\cite{li2021document}. Event detection focuses on identifying trigger words that signify the occurrence of specific events. Event argument extraction seeks to extract the arguments associated with these events from given text. The set of target arguments varies depending on the event's definition.
\begin{itemize}
\item \textbf{Entity Discovery (ED)}~\cite{yan2021unified} encompasses both entity recognition~\cite{cui2021template} and entity typing~\cite{yuan2022generative}. The former is concerned with identifying spans of entities (e.g., ‘Steve’), and the latter focuses on assigning types to these identified entities (e.g., ‘PERSON’).
% % \item \textbf{Relation Extraction (RE)} has various configurations across different wor-ks, primarily categorized into three types: relation classification, relation triplet extraction, and strict relation extraction. Relation classification determines the type of relationship between two given entities. Relation triplet extraction identifies the relation type and the corresponding head and tail entity spans. Strict relation extraction goes further to specify the relation type along with the types of the head and tail entities.
\item \textbf{Event Extraction (EE)}~\cite{qi2023mastering} generally involves two stages. The first stage, event detection~\cite{veyseh2021unleash}, focuses on identifying trigger words that signify the occurrence of specific events. The second stage, event argument extraction~\cite{li2021document}, seeks to extract the arguments associated with these events from given text. The set of target arguments varies depending on the event's definition.
\end{itemize}

Consequently, recent generative IE methods that leverage LLMs to generate structural information have gained more attention than merely extracting it from plain text. Generative IE methods have demonstrated greater flexibility compared to traditional IE approaches. However, due to the limited presence of domain-specific data in LLM training, prompt engineering can be employed to enable LLMs to learn input-output mappings for specific downstream tasks without the need for fine-tuning. Inspired by previous works~\cite{sainz2024gollie, wei2023zeroshot}, our work builds upon a generative IE method to extract information from building rules.

% \textcolor{red}{For better}
% Wei et al.~\cite{wei2023zeroshot} have proposed a unified zero-shot IE model by transforming the zero-shot IE task into a multi-turn question-answering problem with a two-stage framework.

% Given a $n$ tokens input sequence(e.g.,textual rule sequence) $X=[x_1,...,x_n]$, a prompt $P$ and the target information sequence $Y=[y_1,...,y_l]$. Generative IE can be regarded as a form of autoregressive generation. The formula definition of generative IE is as \eqref{FD_GIE} and the objective is to maximize the conditional probability:
% \begin{equation}
% p_{\pi}(Y|X,P)=\prod_{i=1}^{l}p_{\pi}(y_i|y_{<i},X,P)\label{FD_GIE}
% \end{equation}
% where $\pi$ represents the parameters of LLMs, which is frozen in our study,

%\subsection{Code Generation}
\subsection{Code Generation}
\textbf{Code Generation} involves creating programs that adhere to the constraints set by the underlying task~\cite{nijkamp2022codegen}. These constraints typically come in diverse forms, such as input/output pairs, examples, problem descriptions, partial programs, and assertions. The remarkable success of transformers in natural language modeling has sparked significant interest among researchers in leveraging transformer models for code generation.

On one hand, there has been a proliferation of specialized LLMs tailored for code generation. Notably, OpenAI has introduced Codex~\cite{chen2021evaluating}, a GPT-3 model fine-tuned on publicly available code from GitHub, boasting a maximum parameter count of 12 billion. Microsoft unveiled GPT-C~\cite{svyatkovskiy2020intellicode}, a variant of GPT-2 retrained on a vast unsupervised multilingual source code dataset, followed by the release of PyMT5~\cite{clement2020pymt5} and CodeXGLUE~\cite{lu2021codexglue}. On the other hand, rapid engineering techniques are employed to adapt general LLMs for code generation tasks. General LLMs like Llama3~\cite{Llama3} and GPT-4~\cite{GPT4} have also demonstrated impressive performance in various code-related tasks. Additionally, prompt engineering techniques have been proposed to enhance the performance of general LLMs on specific tasks by carefully designing the input prompts.

\section{Methodology}
\label{sec: methodology}
In this section, we introduce ARCEAK, a novel ARC framework, which enhances verification code generation through the integration of architectural knowledge. We begin with an overview of ARCEAK, followed by a detailed explanation of its components in the subsequent subsections.
% {overview:描述任务（以整体设计思路为出发点描述任务）和整体设计；框架简介和图；}
% The establishment of the BIM automatic review system can be divided into four steps, as shown in Fig. 1: first, the structured processing of the rules, second, the establishment of the BIM model and information extraction, third, the execution and reasoning of the review rules, and fourth, the output of the review results7. The first and second steps are the main obstacles to the implementation of automated code checking.

\subsection{Overview}
The implementation of ARC can be delineated into two primary stages: \emph{rule information extraction} and \emph{verification code generation}. The former is dedicated to analyzing the textual rules within the architectural domain and extracting pertinent information essential for verification purposes. The latter stage focuses on the generation of execution code aimed at assessing whether a given architectural blueprint complies with the building rules and requirements.

% The implementation of ARC can be categorized into four stages: building model preparation, rule interpretation, rule checking execution, and checking result reporting. The first and second stages represent the primary challenges in the implementation of ARC. Current ARC research indicates that manual processes are still prevalent, primarily associated with the above tasks.

\begin{figure*}[ht]
% \centerline{\includegraphics{fig/overall_architecture.png}}
\centering
\includegraphics[width=0.9\linewidth]{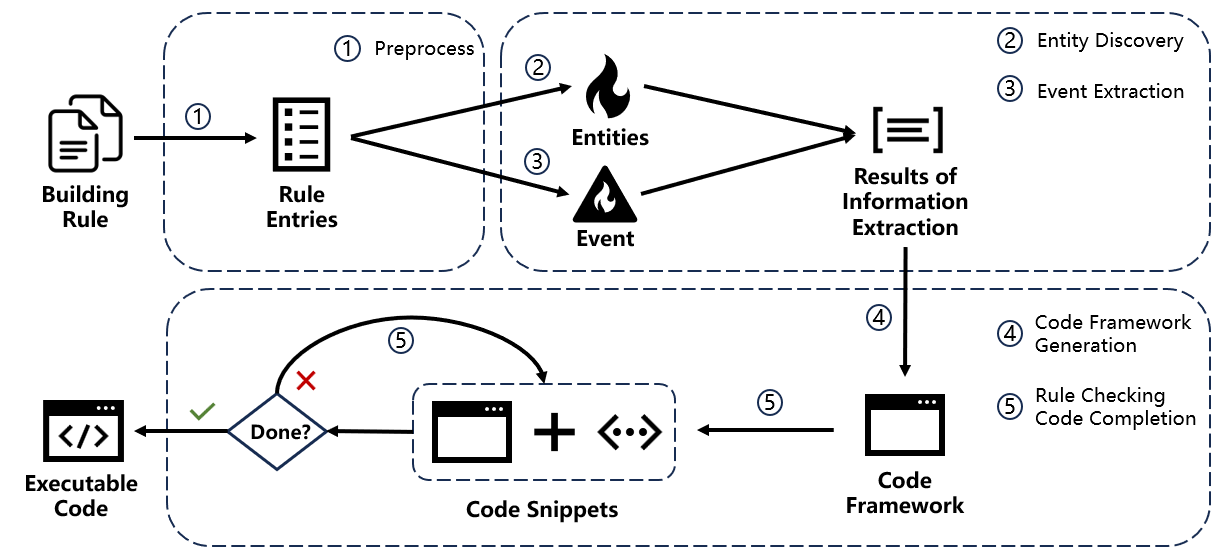}
\caption{The overall architecture of proposed ARCEAK}
\label{overall_architecture}
\end{figure*}

To mitigate or eliminate the necessity for manual annotation and model training during the rule information extraction and verification code generation stages, we introduce the ARCEAK framework, leveraging LLMs. The overall architecture of ARCEAK is illustrated in Fig.~\ref{overall_architecture}. Initially, we undertake a preprocessing phase for building rules, employing a prompt designed for rule splitting and error correction. Subsequently, we proceed to extract pertinent information, such as entities and events, from the refined rule content. Finally, our attention shifts to generating high-level code informed by knowledge-enhanced building rules. The subsequent sub-sections offer a comprehensive elucidation of the distinct phases within ARCEAK.

% To tackle the challenges of time-consuming text annotation, costly pre-training model training, and the oversight of complex semantic information in building rules, we propose ARCEAK, an Automated Rule Checking framework Enhanced with Architectural Knowledge based on large language models. ARCEAK can efficiently extract semantic information, minimize or eliminate the need for manual annotation, enhance building rules with the extracted data, and produce more thorough checking code. The overall architecture of ARCEAK is illustrated in Fig.~\ref{overall_architecture}.

\subsection{Preprocess}

\begin{figure}[htbp]
% \centerline{\includegraphics{fig/pre-processing_result.png}}
\centering
\includegraphics[width=1.0\linewidth]{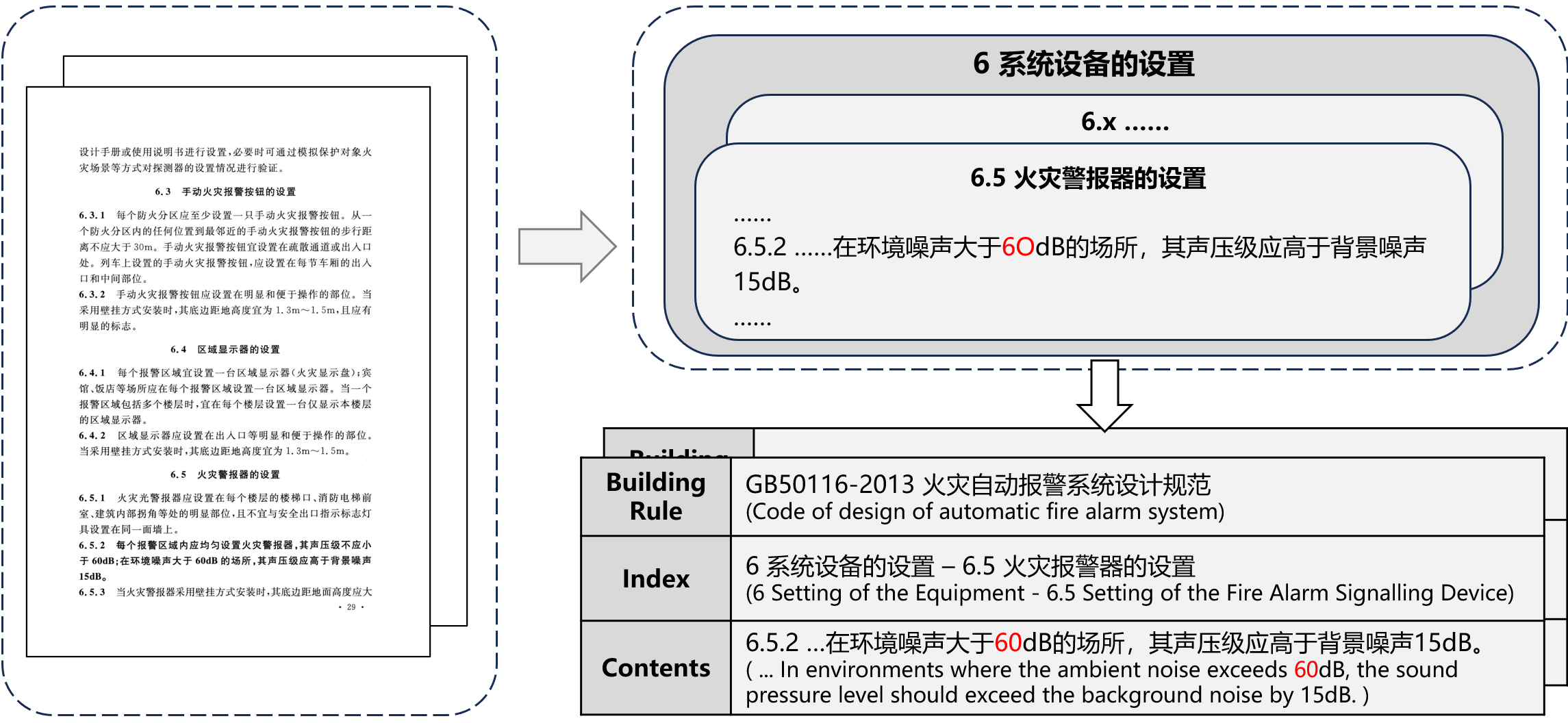}
\caption{An example of preprocessing result}
\label{pre-processing_result}
\end{figure}

Since the national code files are consistently published in PDF format, which is not suitable for information extraction, we need to preprocess the original file and convert it into a structured format. An illustration of the preprocessing results is presented in Fig.~\ref{pre-processing_result}.

First, we convert the PDF file into a TXT file with UTF-8 encoding using Adobe OCR tools. However, the TXT format is not compatible with tables and pictures, which are crucial for understanding the rules and cannot be ignored. After verifying that the number of tables and pictures is manageable, we manually convert the tables into CSV files and the pictures into PNG format.

% Second, we continue preprocessing the text portion of the building rules into a structured format. Typically, building rules are organized into chapters, which are filled with sections, and sections with individual entries. Each rule entry is our main target for information extraction and code generation. Therefore, we utilize an LLM to split the entire chapter into individual rule entries using a few-shot prompting approach. After the initial split, we index each entry with its corresponding chapter and section numbers to serve as a navigational guide. If a rule entry contains tables or pictures, the IDs of these elements are linked to the entry. For example, in the "Code for Design of Automatic Fire Alarm System", we divide it into chapters 1-12. For chapter 6, we further divide it into sections 6.1-6.11. Within section 6.5, we separate it into rule entries 6.5.1-6.5.3. For rule entry 6.5.2, its index is "6 Setting of the Equipment - 6.5 Setting of the Fire Alarm Signalling Device" and its content is "...In environments where the ambient noise exceeds 60dB, the sound pressure level of the alarms should be at least 15dB higher than the background noise."
Second, we continue preprocessing the text portion of the building rules into a structured format. Typically, building rules are organized into chapters, which are filled with sections, and sections with individual entries. Each rule entry is our main target for information extraction and code generation. Therefore, we utilize an LLM to split the entire chapter into individual rule entries using a few-shot prompting approach.
After the initial split, we index each entry with its corresponding chapter and section numbers to serve as a navigational guide. %For instance, in the \textit{Code for Design of Automatic Fire Alarm Systems}, rule 6.5.2 is indexed as '6. Setting of the Equipment - 6.5 Setting of the Fire Alarm Signalling Device', and its content is '... In environments where the ambient noise exceeds 60 dB, the sound pressure level of the alarms should be at least 15 dB higher than the background noise.'

Finally, since similar characters like "0" and "o" or "1" and "I" may be misinterpreted by Adobe OCR tools, leading to semantic errors in subsequent information extraction and code generation stages, OCR mistake detection and correction is required. Additionally, we aim to integrate the content of tables and pictures into the textual content to maintain the alignment of rule entries. Therefore, we leverage the multimodal contextual understanding ability of LLMs by providing the textual content along with the linked tables and pictures to the LLMs, enabling a comprehensive and mistake-free understanding of each rule entry. A manual check is also conducted to ensure the accuracy of the conversion.

\subsection{Rule Information Extraction}
Information Extraction (IE) refers to the process of automatically extracting structured information from unstructured or semi-structured data, such as text. Extracting information from building rules, which encompasses implicit structural entities and intricate constraint conditions, poses several challenges, including complex context understanding and recognition of architectural terminology.

To overcome these issues and extract rule information more accurately and comprehensively, we design an information extraction mechanism based on knowledge-enriched zero-shot prompting with \textbf{C}lassification \textbf{A}nnotation (CA), which requires minimal manual involvement, to extract entities and events from preprocessed rule entries. IE stage mainly contains two tasks: entity discovery and event extraction. %IE mainly contains two tasks: ED and EE.

\begin{table}[htbp]
\caption{Adopted entity types for entity classification}
\begin{center}
\begin{tabular}{cc}
\toprule
\textbf{Type name} & \textbf{Examples} \\
\midrule
% \begin{CJK}{UTF8}{gbsn} 建筑\end{CJK} & \begin{CJK}{UTF8}{gbsn} 建筑, 住宅建筑\end{CJK}\\
Building & building, dwelling building\\
% \begin{CJK}{UTF8}{gbsn} 系统\end{CJK} & \begin{CJK}{UTF8}{gbsn} 灭火系统, 火灾自动报警系统\end{CJK}\\
System & fire extinguishing system,fire automatic alarm system\\
% \begin{CJK}{UTF8}{gbsn} 组件\end{CJK} & \begin{CJK}{UTF8}{gbsn} 探测器, 按钮\end{CJK}\\
Component & detector, button\\
% \begin{CJK}{UTF8}{gbsn} 区域\end{CJK} & \begin{CJK}{UTF8}{gbsn} 防火分区, 防烟分区\end{CJK}\\
Zone & fire compartment, smoke-proof compartment\\
\bottomrule
\end{tabular}
\label{tab:entity types}
\end{center}
\end{table}

% \begin{figure}[htbp]
% \centerline{\includegraphics{IE_structure.png}}
% \caption{The overall workflow of information extraction}
% \label{IE_structure}
% \end{figure}

\textbf{Entity Discovery(ED)} encompasses both entity recognition and entity classification.
% The introduction of LLMs has improved entity discovery, making it effective even with limited labeled data.
To enhance LLMs’ ability to accurately discover domain-specific entities in rule entries, we recommend an adaptive prompt optimization approach in this section. Initially, we allocate instructions for recognizing entities within the given rule entry and concurrently classifying the identified entities into possible types. To address the specificity of entity types, we elaborate on the target entity classification categories to LLMs through entity type classification annotations within the context part of the prompt for the ED task. The CA for ED task may encompass any entity types as long as there are few examples of these types and they perform well with clear and unambiguous heuristic type classification. Here, we employ the four entity types and corresponding examples shown in Table \ref{tab:entity types}.

% \begin{CJK}{UTF8}{gbsn} \end{CJK}

% Moreover, we have implemented formatted handling of entity discovery prediction results by incorporating constraints into the output indicator of the prompt. This enables the generation of responses in JSON format, including the names and types of entities. An example of ED results is illustrated in Table \ref{tab:results of entity discovery and event extraction}.

% \textbf{Event Extraction(EE)} involves identifying and classifying events described in text. An event, serving as a fine-grained semantic unit to describe the state of entities and their actions, is typically defined as a textual span comprising a predicate and its arguments. Complex conditions pose one of the primary challenges in interpreting building rules. Current research on ARC for rules with conditions still necessitates significant manual effort in entity labeling or sentence reconstruction \cite{zhou2022integrating, peng2023automated}. Rule entries in building rules can be abstracted into two categories: attribute assignment with conditions and attribute assignment without conditions.
\textbf{Event Extraction(EE)} involves identifying and classifying events described in text. An event, serving as a fine-grained semantic unit to describe the state of entities and their actions, is typically defined as a textual span comprising a predicate and its arguments. Complex conditions pose one of the primary challenges in interpreting building rules. Rule entries in building rules can be abstracted into two categories: attribute assignment with conditions and attribute assignment without conditions.

In our study, we define an attribute assignment as a type of event, where entities identified in the ED phase serve as ``trigger words'', with the goal of extracting arguments related to "assignment" events. A complete argument entry consists of the following components:
\begin{itemize}
\item \textbf{Entity of Attribute:} The entity to which the attribute belongs.
\item \textbf{Attribute Name:} The name of the extracted attribute.
\item \textbf{Conditions or Constraints:} The constraints on attribute assignment.
\item \textbf{Comparator:} The comparator used for the attribute value.
\item \textbf{Attribute Value:} The value assigned to the attribute.
\end{itemize}

% The arguments, which describe the core information of "assignment" events, are quite complex for LLMs to directly identify and extract without further explanation. To enhance the external knowledge of LLMs, we incorporate CA into the context part of the prompt for the EE task. The CA for the EE task defines six types of attributes as the target for LLMs to classify, employing a heuristic approach.
% This CA framework, designed specifically for the EE task, outlines six distinct types of attributes that LLM is required to classify. By integrating CA into the task, LLM is provided with a heuristic approach to effectively identify and extract the core information needed for successful completion of the assignment. Like in the ED task, we also guide the LLM on the desired format of response by incorporating event extraction response format into the output indicator of the prompt. An example of EE results is illustrated in Table \ref{tab:results of entity discovery and event extraction}.

Since general LLMs lack domain-specific knowledge of construction, the arguments describing the core information of 'assignment' events present considerable complexity, making it difficult for LLMs to identify and extract them directly without additional explanation. To enhance the LLMs' ability to recognize assignments in the construction domain, we compiled common assignment expressions for six types of assignments. The examples of common assignment expressions are shown in Table \ref{tab:expression of assignment}.

\begin{table}[htbp]
\caption{Examples of common expressions of assignments}
\begin{center}
\begin{tabular}{cc}
\toprule
\textbf{Type name} & \textbf{Examples} \\
\midrule
% \begin{CJK}{UTF8}{gbsn} 直接属性约束 \end{CJK} & \begin{CJK}{UTF8}{gbsn} 长度，宽度，高度\end{CJK}\\
Direct attribute constraint & length, width, height\\
% \begin{CJK}{UTF8}{gbsn} 数量约束\end{CJK} & \begin{CJK}{UTF8}{gbsn} 只，个，扣\end{CJK}\\
Quantity constraint & quantity, piece, unit\\
% \begin{CJK}{UTF8}{gbsn} 距离约束\end{CJK} & \begin{CJK}{UTF8}{gbsn} 距，距离，相隔\end{CJK}\\
Distance constraint & distance, spacing, separation\\
% \begin{CJK}{UTF8}{gbsn} 分类约束\end{CJK} & \begin{CJK}{UTF8}{gbsn} 类，类别\end{CJK}\\
Classification constraint & type, category\\
% \begin{CJK}{UTF8}{gbsn} 空间约束\end{CJK} & \begin{CJK}{UTF8}{gbsn} 顶部，上方，居中\end{CJK}\\
Spatial constraint & top, above, centered\\
% \begin{CJK}{UTF8}{gbsn} 其他间接约束\end{CJK} & \begin{CJK}{UTF8}{gbsn} 有组件，由组成\end{CJK}\\
Other indirect constraint & components, composed of\\
\bottomrule
\end{tabular}
\label{tab:expression of assignment}
\end{center}
\end{table}

% To enrich the external knowledge of LLMs, we integrate CA into the context part of the prompt for the EE task. This CA, specifically designed for the EE task, delineates six distinct types of attributes that LLMs are required to classify using a heuristic approach. By incorporating CA into the task, LLMs are provided with a heuristic approach to effectively identify and extract the core information necessary for successful assignment completion. Similar to the ED task, we guide the LLM on the desired response format by incorporating the EE result response format into the output indicator of the prompt.

After the IE stage, the entities and events in the construction domain are extracted from the rule entry content, and the unstructured rule entry content is formatted as structured JSON lists.

\subsection{Verification Code Generation}
% LLMs, which trained on huge amounts of natural language description and code pair, have the ability to generate code for common tasks(e.g., algorithm solving, software developing),
Code generation refers to the process of automatically generating source code by a computer program. Manual code writing and verification is expensive, and generating code from structured rules is effective for explicit rules but often incomplete due to the presence of implicit rules. To address this, we propose a knowledge-augmented code generation workflow that leverages the code generation and completion capabilities of LLMs. Given the complexity of verification code generation, we follow the approach of Plan-and-Solve Prompting~\cite{DBLP:conf/acl/WangXLHLLL23}, dividing the entire process into two primary steps: code framework generation and rule-checking code completion.

% After the information extraction phase, structured aggregated information is extracted from the implicitly loose descriptions in the rule entries. Although this information is structured as natural language descriptions, it is not in a format suitable for ARC. To assist LLMs in transforming these building rules into executable code more comprehensively and accurately, we propose a multi-step code generation workflow based on zero-shot code generation prompts.

\smallskip
% \noindent
\textbf{Code Framework Generation} aims to construct a fundamental code skeleton, encompassing vital sections for variable initialization, function definitions, and control structures necessary for conditional evaluations. To enable LLMs to grasp the domain knowledge of building rules and generate precise code representing the details of rule entries, as well as to control the granularity of the function code and align the code structure more closely with the specific requirements of the building rules, we integrate entities and arguments related to the "assignment" event extracted during the IE stage with code framework generation prompt instructions. To address semantic dependencies between rules, we use a parser to determine whether the current rule entry depends on other rule entries; if such dependencies exist, both the current rule entry and its dependencies are provided to the LLM. Furthermore, to enhance the coverage rate, we aim to guide LLMs to annotate the generated verification code with the target rule entry index, ensuring that each rule component presented in the rule entries is comprehensively checked and represented in the code framework. Our code generation prompt is shown in Table \ref{tab:Code Generation Prompt}.

\begin{table}[htbp]
    \caption{Code generation prompt}
    \begin{center}
    \begin{tabular}{c}
        \toprule
        \textbf{Code Generation Prompts} \\
        \midrule
        \parbox[c]{8.3cm} {\textbf{Instruction: }
        % \begin{CJK}{UTF8}{gbsn}你是一个revit二次开发工程师，你擅长编写revit合规性检查代码，请将``` '''包围的建筑规范转换成revit检查代码，请以C\#语言编写，请参考从规范中提取的实体和属性，编写合规性检查代码。\end{CJK}
        You are a Revit secondary development engineer who is skilled in writing Revit compliance check code. Please convert the building specifications surrounded by ``` ''' into Revit check code in C\# language. Please refer to the entities and properties extracted from the specifications and write compliance check codes.} \\
        % \parbox[c]{12cm}{\textbf{Input Data: }$\langle$entity list in JSON format$\rangle$, $\langle$event list in JSON format$\rangle$, '''$\langle$rule content$\rangle$'''}\\\\
        \parbox[c]{8.3cm}{\textbf{Input Data: }$\langle$entity list$\rangle$, $\langle$event list$\rangle$, '''$\langle$rule content$\rangle$'''}\\
        % \multirow{3}{*}{Input data}  & \begin{CJK}{UTF8}{gbsn}实体列表：\end{CJK}$\langle$ entity list$\rangle$ \\
        %                              & \begin{CJK}{UTF8}{gbsn}事件列表：\end{CJK}$\langle$ Event list $\rangle$ \\
        %                              & \begin{CJK}{UTF8}{gbsn}规则：\end{CJK}   $\langle$ Rules $\rangle$      \\
        \parbox[c]{8.3cm}{\textbf{Output Format: }Please first generate the code framework. The framework should include basic structures for variable initialization, function definitions, and conditional statements. For unimplemented functions, please add $\langle$unimplemented$\rangle$ before the function definition. Ensure that each specification check item appears in the code.} \\
        \bottomrule
    \end{tabular}
    \label{tab:Code Generation Prompt}
    \end{center}
\end{table}

\textbf{Rule-Checking Code Completion} begins with integrating detailed logic and specific code snippets into the basic code skeleton formed during code framework generation. This crucial phase involves populating the established framework with specific details and logic tailored to the building rules. Placeholders previously defined in the skeleton are filled with actual code sections generated to check compliance with the rules. Furthermore, to pinpoint the specific rule that an architectural blueprint violates, we instruct LLMs to add detailed assert statements, including the specific rule entry according to the indexed comments in the code. Our code completion prompt is shown in Table \ref{tab:Code Completion Prompt}.

\begin{table}[htbp]
    \caption{Code completion prompt}
    \begin{center}
    \begin{tabular}{cc}
        \toprule
        \textbf{Code Completion Prompts} \\
        \midrule
        \parbox[c]{8.3cm} {\textbf{Instruction: }You are a Revit secondary development engineer who is skilled in writing Revit compliance check code. Please convert the building specifications surrounded by ``` ''' into Revit check code in C\# language. Please refer to the entities and properties extracted from the specifications and write compliance check codes. We will provide a code framework. Please implement the unimplemented functions in the code framework enclosed by ''' '''}\\
        % \parbox[c]{12cm}{\textbf{Input Data: }$\langle$code framework$\rangle$,$\langle$unimplemented function$\rangle$, $\langle$entity list in JSON format$\rangle$, $\langle$event list in JSON format$\rangle$, ```$\langle$rule content$\rangle$'''}\\\\
        \parbox[c]{8.3cm}{\textbf{Input Data: }$\langle$code framework$\rangle$,$\langle$unimplemented function$\rangle$, $\langle$entity list$\rangle$, $\langle$event list$\rangle$, ```$\langle$rule content$\rangle$'''}\\
        \parbox[c]{8.3cm}{\textbf{Output Format: }Please provide the complete code.}\\
        \bottomrule
    \end{tabular}
    \label{tab:Code Completion Prompt}
    \end{center}
\end{table}

Since new unimplemented or undefined functions may be added during the code completion process, iterations are necessary. To determine if the code completion is complete, we design a parser to extract all variables and functions used in the generated verification code, then verify whether each variable and function is defined and implemented. Additionally, because it is uncertain whether the code generated by LLMs is runnable and may contain compilation errors, we employ a code self-refinement process to enhance the runnability of the generated verification code. In this process, we provide the LLMs with the completed verification code along with reported errors, allowing them to refine the code they generated.

\begin{figure*}[htpb]
% \centerline{\includegraphics{fig/pre-processing_result.png}}
\centering
\includegraphics[width=1.0\linewidth]{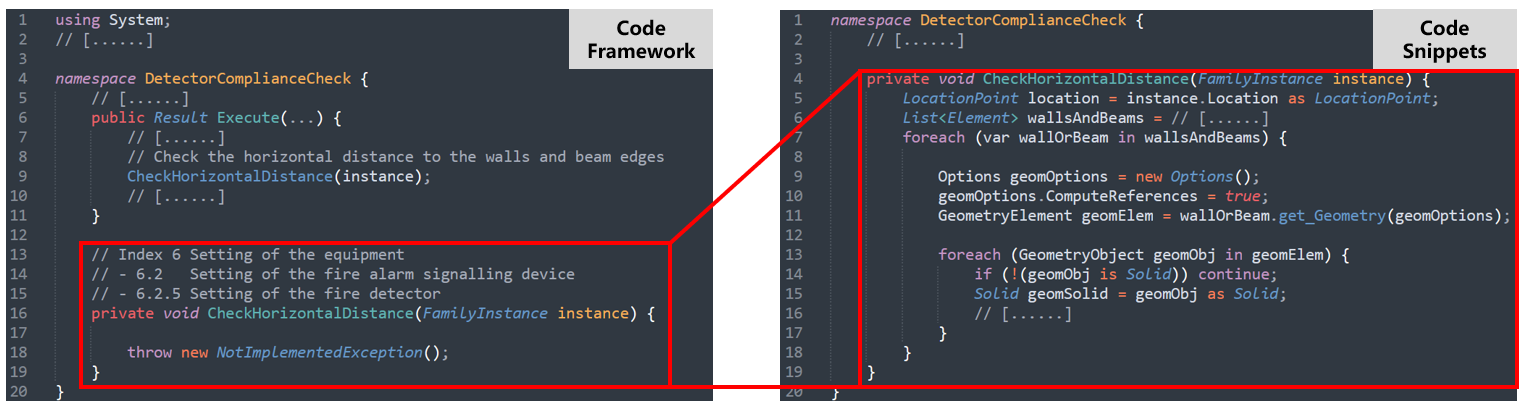}
\caption{An example of code generation result}
\label{code_generation}
\end{figure*}

After the knowledge-augmented verification code generation stage, we facilitate a smoother transition from natural language descriptions of building rules to executable verification code. This code can be used to verify whether architectural blueprints comply with regulations by invoking the API of the selected models. An example of code generation result is shown in Fig. \ref{code_generation}.

% This phase often necessitates several iterations of code generation to cover various rule aspects, including exception handling, condition validation, and rule-specific actions. Each iteration is subjected to thorough refinement to ensure that the code not only functions accurately but also efficiently enforces the rules.

% By employing this structured, multi-step workflow, we facilitate a smoother transition from natural language descriptions to executable code, enhancing the ability of LLMs to automate the enforcement of complex building rules both accurately and comprehensively.

\section{Evaluation}
\label{sec: evaluation}
% For building rules data, we choose the Chinese building code \textit{GB50116-2013} \begin{CJK}{UTF8}{gbsn}火灾自动报系统设计规范\end{CJK} (Code for Design of Automatic Fire Alarm System) because the Fire Alarm System Code is one of the most important regulations in the construction domain. Additionally, its semantic information is very clear, meaning that most of its rules are mandatory rather than advisory. Furthermore, we select three room-level BIM (Building Information Modeling) models as the target models for verification code generation, as BIM is one of the most widely-used construction models in the world.

% For the LLMs, we use GPT-3.5-Turbo, published by OpenAI \footnote{https://openai.com/}, for Rule Information Extraction stage and we implement our  with both GPT-3.5-Turbo and GPT-4-Turbo,

To comprehensively evaluate the performance of the proposed framework ARCEAK, we conduct a large-scale study to seek to address the following research questions (RQ):
\begin{itemize}
    \item \textbf{RQ1: How effective is ARCEAK in extracting information from a building rule?} As described above, we employ a knowledge-enriched zero-shot prompting strategy to enhance LLM performance in the IE stage. This RQ aims to verify whether the incorporation of CA in ARCEAK improves the LLM's ability to perform IE more effectively.
    \item \textbf{RQ2: How comprehensive and accurate is the code generated by ARCEAK in enforcing building rules? } We employed a two-step code generation process, enhanced by knowledge extracted during the IE stage. This RQ seeks to determine whether the knowledge augmentation and two-step generation approach in ARCEAK positively impact domain-specific code generation.
    \item \textbf{RQ3: How does the IE stage in ARCEAK enhance the accuracy and efficiency of verification code generation stage?} ARC is a domain-specific and complex task, and merely providing evaluation metrics may not be sufficient to fully convey the effectiveness of ARCEAK. It is valuable to analyze concrete cases to better understand the impact of ARCEAK's IE stage on generating building compliance checking code.

% LLM是否有助于实体挖掘ED EE
% 第二个phase
% 第一部分的结果是否有助于第二个部分能否运行、覆盖率、testcase(正负样本)
\end{itemize}
% \subsection{Research Question}

% The building rule GB50116 is one of the core building rule in building electricity,
% non-compliances of

% \begin{itemize}
    % \item \textbf{Dataset.} In our study, the Chinese building code \textit{GB50116-2013}(Code for Design of Automatic Fire Alarm System) is selected to validate the IE phase in ARCEAK. Additionally, to verify the accuracy of the code generation results, we adopted three room-level BIM (Building Information Modeling) models. These models contain components that both comply with and deviate from the selected rule.
    % \item \textbf{Experiment Setup.} For the IE phase of ARCEAK, we implement our methodology using GPT-3.5\footnote{https://openai.com/}, and compare it to CoT prompt without CA, which we call naive CoT. This comparison is conducted across both ED and EE tasks. In the naive baseline, the model is given the natural language instruction and is asked to directly discover the entity and extract events related to the entity. For code generation phase of ARCEAK, we implement our method with GPT-3.5 and GPT-4 \footnote{https://openai.com/}, and compare to code generation without knowledge-augmention.

% \end{itemize}

% api 版本
\noindent \textbf{RQ1:How effective is ARCEAK in extracting information from a building rule? }

\smallskip\noindent\textbf{Setup.} In our study, the Chinese building code \textit{GB50116-2013} (Code for Design of Automatic Fire Alarm System) is selected to validate the IE stage in ARCEAK. For the IE stage of ARCEAK, we implement our method using GPT-3.5\footnote{https://openai.com/}, and compare it to Chain-of-Thought(CoT) Prompting~\cite{wei2022chain} without CA, which we call naive CoT. This comparison is conducted across both ED and EE tasks. In the naive baseline, the model is given the natural language instruction and is asked to directly discover the entity and extract events related to the discovered entity.

% TP denotes correct predictions made by the classifier for different entity types, indicating that the predicted entity type aligns with the actual one
% precision, which is calculated as the number of correct positive predictions (TP) divided by the total number of positive predictions (TP + FP) made by the model, and recall, which is the number of true positives (TP) divided by the total number of positive instances (TP + FN). Precision reflects the reliability of the model when it predicts an instance as positive, whereas recall indicates the model's ability to identify all positive instances within the dataset.

\smallskip\noindent\textbf{Metrics.} To compare the performance of the ED phase of IE stage, we employed the following three distinct metrics. Prior to introducing these metrics, it is essential to revisit three fundamental concepts: True Positives(TP), False Positives(FP) and False Negatives(FN). TP denotes correct predictions made by the classifier for different entity types. Conversely, FP and FN denote instances from different type of entities that have been misclassified.
\begin{itemize}
    \item \textit{Precision}: Precision is the ratio of the number of entities extracted from rule with correct types to the number of entities extracted from rule
    , which is calculated with,
\begin{equation}
Precision = \frac{TP}{TP + FP}\label{Precision}
\end{equation}
    \item \textit{Recall}: Recall is the ratio of the number of entities extracted from rule with correct types to the number of entities which should be extracted from rule(ground truth)d
    , which is calculated with,
\begin{equation}
Recall = \frac{TP}{TP + FN}\label{Recall}
\end{equation}
\item \textit{F1}: F1 score is a weighted average of the framework precision and recall, which ranges from 0 to 1. A higher F1 score indicates better comprehensive performance of the framework. F1 score is defined as the harmonic mean of the precision and recall, which can be calculated with,
% $F1 = 2 \cdot \frac{Precision\cdot Recall}{Precision + Recall}$.
\begin{equation}
F1 = 2 \cdot \frac{Precision\cdot Recall}{Precision + Recall}\label{F1}
\end{equation}
\end{itemize}
To comprehensively evaluate the performance of the EE phase of IE stage, we employed four distinct metrics,
\begin{itemize}
    \item \textit{Tri-R}: Tri-R is the ratio of the number of intersection of events extracted from rule and events which should be extracted from rule with the same "trigger word" entity to the number of events which should be extracted from rule(ground truth), which represents the ratio of how many real events are extracted.
    % Tri-R is calculated as \eqref{Recall}.
    \item \textit{Arg-P$_{a}$}: Arg-P$_{a}$ is the ratio of the number of intersection of events extracted from rule and ground truth of event with the same entity and attribute to the number of events extracted from rule, which represents the ratio of how many extracted events extract the true attribute.
    % Arg-P$_{a}$ is calculated as \eqref{Precision}.
    \item \textit{Arg-P$_{A}$}: Arg-P$_{A}$ represents the ratio of extracted events that contain all the required components.
    % Arg-P$_{A}$ is calculated as \eqref{Precision}, where TP is the number of events contain all true arguments(attribute name, conditions or constraints, comparator and attribute value).
    \item \textit{Arg-P$_{O}$}: Arg-P$_{O}$ represents the ratio of extracted events that contain all the required components and are returned in the correct order as specified by the prompt.
\end{itemize}
% \begin{itemize}
% \textbf{For IE phase:} To comprehensively compare the performance of the IE phase of ARCEAK, we considered seven different metrics. Before introducing these metrics, some concepts need to be defined in advance. The variables representing these concepts are shown in Table \ref{tab:variable descriptions}.
% \begin{table}[htbp]
% \caption{Variable Names and Descriptions}
% \begin{center}
% \begin{tabular}{cc}
% \toprule
% \textbf{Variable Name} & \textbf{Description} \\
% \midrule
% $E_{G}$ & \parbox[c]{6cm}{\centering entities to be identified and extracted.} \\
% $E_{e}$ & \parbox[c]{6cm}{\centering entities extracted from rule} \\
% $E_{e_r}$ & \parbox[c]{6cm}{\centering $E_{G} \cap E_{E}$} \\
% $E_{C}$ & \parbox[c]{6cm}{\centering $E_{e_r}$ with correct type} \\
% $A_{G}$ & \parbox[c]{6cm}{ \centering Events to be identified and extracted} \\
% $A_{E}$ & \parbox[c]{6cm}{\centering events extracted from rule} \\
% $A_{E_r}$ & \parbox[c]{6cm}{\centering $A_{G} \cap A_{AE}$} \\
% $A_{E_{wea}}$ & \parbox[c]{6cm}{\centering events contain correct entity and attribute} \\
% % $A_{E_{weac}}$ & \parbox[c]{6cm}{\centering $A_{wea}$ with correct comparator} \\
% $A_{wA}$ & \parbox[c]{6cm}{\centering $A_{wea}$ contain all arguments} \\
% $A_{O}$ & \parbox[c]{6cm}{\centering $A_{wA}$ with correct order} \\
% \bottomrule
% \end{tabular}
% \label{tab:variable descriptions}
% \end{center}
% \end{table}

\begin{table}[htbp]
\caption{Evaluation on classification annotation in entity discovery stage}
\begin{center}
\begin{tabular}{c|cc}
\toprule
& \textbf{Naive CoT} & \textbf{ARCEAK} \\
\midrule
\textbf{Number of Extracted Entities} & 407 & \textbf{774} \\
\textbf{Precision} &  0.042& \textbf{0.631} \\
\textbf{Recall} &  0.020& \textbf{0.713} \\
\textbf{F1} &  0.027& \textbf{0.669}\\
\bottomrule
\end{tabular}
\label{tab:evaluation on CA in ED}
\end{center}
\end{table}

\smallskip\noindent\textbf{Result and Analysis.}
% All of the evaluator and architecture experts are from East China Architectural Design \& Research Institute.
The primary goal of RQ1 is to assess the effectiveness of CA in ARCEAK for extracting information with minimal manual involvement. During the ED phase of the IE stage, the LLM extracts and classifies potential architectural entities. In the EE phase, the LLM detects ``assignment'' events based on ``trigger word'' entities and extracts their potential arguments. These experiments use the entire set of rule entries from GB50116-2013. To ensure accurate evaluation, a two-layer assessment structure is implemented. The first layer involves five junior evaluators cross-evaluating the IE results, with each result reviewed by two evaluators. In the second layer, a domain expert equipped with empirical knowledge further assesses the results.

Table \ref{tab:evaluation on CA in ED} shows ARCEAK's performance in the ED phase of the IE stage. ARCEAK extract 90\% more entities than naive CoT prompting, while achieving approximately 64\% higher F1 score. The data indicates that CA significantly enhances the LLM's ability to extract and classify architectural entities. The poor performance of the naive CoT approach, lacking CA, highlights the limitations of LLMs in extracting domain-specific information. In contrast, the improved performance with CA underscores its effectiveness in the ED phase.

\begin{table}[htbp]
% \caption{Evaluation on Classification Annotation(CA) in event extraction stage of ARCEAK}
\caption{Evaluation on classification annotation in event extraction stage}
\begin{center}
\begin{tabular}{c|cc}
\toprule
& \textbf{Naive CoT} & \textbf{ARCEAK} \\
\midrule
\textbf{Number of Extracted Event} & 659 & \textbf{693} \\
\textbf{Tri-R} &  0.839& \textbf{0.844} \\
\textbf{Arg-P$_a$} &  0.553& \textbf{0.535} \\
\textbf{Arg-P$_A$} &  0.378& \textbf{0.365}\\
\textbf{Arg-P$_O$} &  0.205& \textbf{0.227}\\
\bottomrule
\end{tabular}
\label{tab:evaluation on CA in EE}
\end{center}
\end{table}
Table \ref{tab:evaluation on CA in EE} shows the comprehensive performance of ARCEAK in the EE phase of the IE stage, which demonstrates that prompting with CA can improve the event detection and argument extraction capabilities of the LLM. In the EE phase, CA resulted in 1.8\% and 1.3\% lower performance in Arg-P$_a$ and Arg-P$_A$, respectively. However, it achieved a 2.2\% higher performance in Arg-P$_O$. This indicates that while the LLM prompted by naive CoT tends to retain more information in the extracted events without fully understanding the relevance of the retained information, CA in the EE phase enhances the precision of extracting specific types of arguments.

\begin{tcolorbox}[colback=gray!10,
                  boxrule=.2mm,
                  arc=2mm,
                  auto outer arc,
                  before upper={\parindent15pt\noindent},
                  ]
    \textbf{Answer to RQ1:} ARCEAK significantly improves the precision, recall, and F1 score in the ED phase. In the EE phase, ARCEAK enhances the LLM's ability to detect events and extract arguments with higher precision for specific argument types. Overall, ARCEAK's CA method effectively enhances the LLM's performance in IE stage.
    % \textbf{Answer to RQ1:} ARCEAK significantly improves the precision, recall, and F1 score in the ED stage, demonstrating its ability to accurately extract and classify architectural entities. In the EE stage, ARCEAK enhances the LLM's ability to detect events and extract arguments with higher precision for specific argument types, although some slight trade-offs are observed in overall argument precision. ARCEAK's Classification Annotation method effectively enhances the LLM's performance in information extraction tasks.
\end{tcolorbox}

\noindent\textbf{RQ2: How comprehensive and accurate is the code generated by ARCEAK in enforcing building rules?}\smallskip

\smallskip\noindent\textbf{Setup.} For verification code generation stage of ARCEAK, rule entries from GB50116-2013 are selected and twenty room-level BIM  models containing components that both comply with and deviate from the selected rule are constructed to evaluate the performance of prompt with CoT. We implement our method with GPT-3.5 and GPT-4, and compare to code generation without CoT prompt. The code generated by LLM is running on Revit 2020 to test for the accurate performance.
% We generate the selected rule entries(excluding those with no corresponding component in )

\smallskip\noindent\textbf{Metrics.} To comprehensively analysis the performance and cost of CoT prompt in code generation phase, we introduce the following metrics:

% \textit{Code Integrity}: Code Integrity refers to the extent to which the generated code framework during the initial phase accurately represents and retains the requirements outlined in the building code.
\begin{itemize}
    \item \textit{Code Integrity}: Code Integrity refers to the extent to which the generated code framework during the initial phase accurately represents and retains the requirements outlined in the building code.
    \item \textit{Done@K}: Done@K is the ratio of the number of code completions that are successfully finalized before the K+1 term to the total number of code generations.
    \item \textit{Compile Pass Rate}: Compile Pass Rate is the ratio of the number of the generated codes which raise no error while compiling to the number of the generated codes.
    \item \textit{Logic Pass Rate}: Logic Pass Rate is the ratio of the number of the generated codes which contain no logic error(e.g., wrong comparison) to the number of the generated codes.
    \item \textit{Pass Rate}: Pass Rate is the ratio of the number of the generated codes pass the test model to the number of the generated codes.
\end{itemize}

\begin{table}[htbp]
\caption{Comparison between GPT-3.5 and GPT-4 on code integrity and Done@K}
\begin{center}
\begin{tabular}{c|cccc}
\toprule
& \multicolumn{2}{c}{\textbf{wo-CoT}} & \multicolumn{2}{c}{\textbf{w-CoT}}\\
& \textbf{GPT-3.5} & \textbf{GPT-4} & \textbf{GPT-3.5} & \textbf{GPT-4} \\
\midrule
\textbf{Code Integrity}  & 0.78   & 0.93 & 1.00 & 1.00 \\
\textbf{Done@1}          & 0.36   & 0.30 & 0.20 & 0.24 \\
\textbf{Done@2}          & 0.94   & 0.90 & 0.90 & 0.88 \\

% \textbf{Check }
% \textbf{Runable Rate} & 0\% & 24.77\% & 24.89\% & 17.12\% \\
\bottomrule
\end{tabular}
\label{tab:Comparasion between GPT-3.5 and GPT-4 on Code Integerity and Done@K}
\end{center}
\end{table}

\smallskip\noindent\textbf{Result and Analysis.} The RQ2 is mainly to evaluate whether the CoT prompt in ARCEAK for code generation is effective on generating codes that closely align with the specific requirements of rule entries. The CoT prompt is employed to enhance the LLM's ability to interpret and implement detailed rule entries effectively. By using CoT, the LLM is prompted to reason through each step or requirement in a rule entry before generating the corresponding code.

Table \ref{tab:Comparasion between GPT-3.5 and GPT-4 on Code Integerity and Done@K} represents the performance of ARCEAK in aligning code with rule entries accurately. With the introduction of CoT in LLM, the code integrity is improved to 100\%, which means no part of the rule is overlooked and that the final output adheres closely to the specified requirements. Due to the instruction of a two-section generation strategy(generate the code framework first and then complete the unimplemented functions), Done@1 is decreased to approximately 20\%, which declines potentially escalates the computational and temporal costs associated with simpler rule entries.
Nevertheless, Done@2 metric of code generation prompt with CoT achieves similar rates compared to the prompt without CoT, suggests that the model can effectively adjust and improve its initial outputs. This approach is particularly beneficial for complex coding tasks, where the initial framework helps ensure more precise completions in subsequent steps.

\begin{table}[htbp]
\caption{Comparasion between GPT-3.5 and GPT-4 on 0-shot and 1-shot code framework generation}
\begin{center}
\begin{tabular}{c|cccc}
\toprule
& \multicolumn{2}{c}{\textbf{0-shot}} & \multicolumn{2}{c}{\textbf{1-shot}}\\
\cmidrule(lr){2-3}\cmidrule(lr){4-5}
& \textbf{GPT-3.5} & \textbf{GPT-4} & \textbf{GPT-3.5} & \textbf{GPT-4} \\
\midrule
\textbf{Compile Pass} & 0.050   & 0.250  & 0.630  & 0.580 \\
\textbf{Logic Pass}   & 0.020   & 0.130  & 0.110  & 0.240 \\
\textbf{Pass }        & 0.0     & 0.080  & 0.030  & 0.100 \\
% \textbf{Check }
% \textbf{Runable Rate} & 0\% & 24.77\% & 24.89\% & 17.12\% \\
\bottomrule
\end{tabular}
\label{tab:Comparasion between GPT-3.5 and GPT-4 on zero and few-shot code framework generation}
\end{center}
\end{table}

Table \ref{tab:Comparasion between GPT-3.5 and GPT-4 on zero and few-shot code framework generation} compares the performance of GPT-3.5 and GPT-4 in both 0-shot and 1-shot code framework generation scenarios. In the 1-shot scenario, GPT-3.5 exhibits a higher compile pass rate at 63\% compared to GPT-4's 58\%. This higher rate is attributed to GPT-3.5's tendency to use annotations or custom variables instead of direct component and property access operations, which reduces compilation errors. However, this approach leads to a significant drop in the logic pass rate and overall pass rate for GPT-3.5 compared to GPT-4. This is because the code generated by GPT-3.5 often fails to access model parameters correctly, affecting the functionality and logic accuracy of the code.

% Nevertheless, the stabilization of performance at the Done@2 metric, where it achieves similar rates compared to the prompt without CoT, suggests an effective recovery mechanism inherent in the LLM's design. Such a mechanism is particularly advantageous for intricate coding tasks, as the foundational framework established in the initial step significantly contributes to the accuracy of subsequent completions. This dual-phase generation approach, despite its initial decrement in efficiency, demonstrates the model’s capacity to adjust and refine outputs, thereby aligning closely with the detailed requirements of rule entries on a second attempt.

% \begin{table}[htbp]
% \caption{Evaluation on Code Generation stage of ARCEAK}
% \begin{center}
% \begin{tabular}{ccccc}
% \toprule
% & \multicolumn{2}{c}{\textbf{GPT-3.5}} & \multicolumn{2}{c}{\textbf{GPT-4}}\\
% & \textbf{0-shot} & \textbf{1-shot} & \textbf{0-shot} & \textbf{1-shot} \\
% \midrule
% \textbf{Compile Pass} & 2.5\% & 70\% & 20\% & 45,894 \\
% \textbf{Logic Error} & 15.98\% & 18.46\% & 18.58\% & 11.30\% \\
% \textbf{Pass} & 33.70\% & 37.65\% & 37.65\% & 35.26\%\\
% \textbf{Runable Rate} & 21.68\% & 24.77\% & 24.89\% & 17.12\% \\
% \bottomrule
% \end{tabular}
% \label{tab:Comparasion between GPT-3.5 and GPT-4 on zero and few-shot code framework generation}
% \end{center}
% \end{table}

\begin{tcolorbox}[colback=gray!10,
                  boxrule=.2mm,
                  arc=2mm,
                  auto outer arc,
                  before upper={\parindent15pt\noindent},
                  ]
    \textbf{Answer to RQ2:} The code generated by ARCEAK, particularly when using advanced models like GPT-4, is fairly comprehensive and accurate in enforcing building rules, despite some initial inefficiencies in the 1-shot scenario. The model's ability to adjust and refine its output ensures that it can meet detailed rule requirements effectively, making it suitable for complex tasks that demand high precision and adherence to specific standards.
\end{tcolorbox}

\noindent\textbf{RQ3: How does the IE stage in ARCEAK enhance the accuracy and efficiency of verification code generation stage?}\smallskip

\smallskip\noindent\textbf{Setup.}
To answer the RQ3, we implement our method with GPT-3.5 and GPT-4, and compare to code generation without the assistance of knowledge augmentation, which serves as a baseline to assess the effect of the IE stage.

\smallskip\noindent\textbf{Case Study.}
The RQ3 is proposed for analyzing the impact of knowledge augmentation on verification code generation and how extracted information from rule entries contributes to improved code accuracy and efficiency. The following case studies focus on specific instances where knowledge augmentation has markedly influenced the output of the code generation process. By examining these instances, we aim to illustrate concrete examples of both success and challenges in integrating extracted information, providing a deeper understanding of the mechanisms and factors that influence the outcomes.
% \begin{itemize}
%     \item \textbf{Enhancement on compatibility between the LLMs and selected model}:
% \end{itemize}
\begin{figure}[htpb]
\centering
\includegraphics[width=1.0\linewidth]{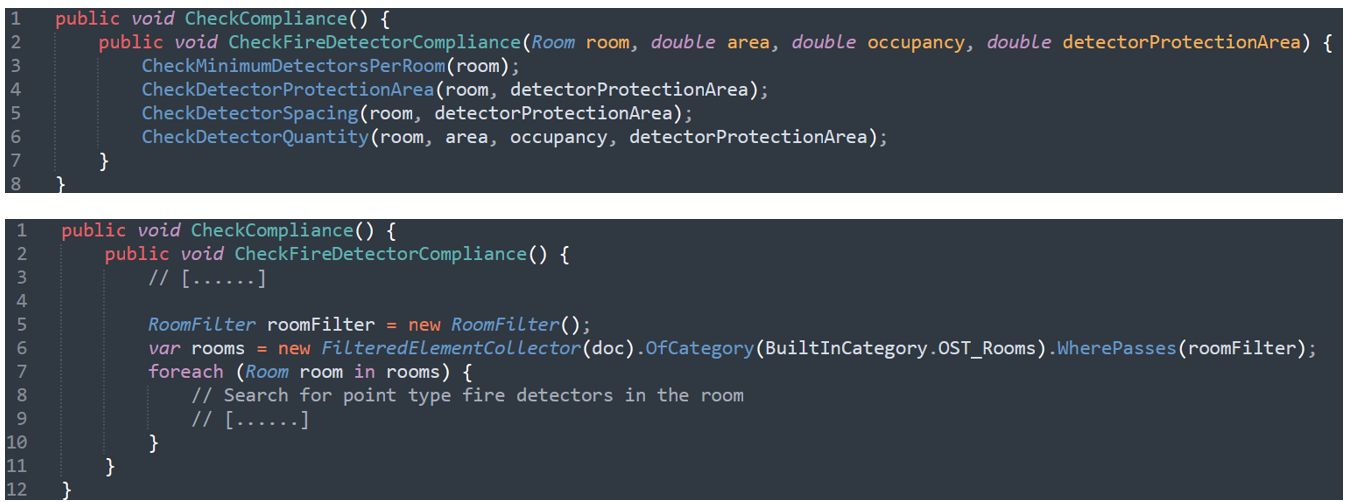}
\caption{An example on compatibility between the LLMs and selected model. The code above is generated without external information and the code below is generated with entity information extracted in Entity Discovery.}
\label{RQ3code1}
\end{figure}
\begin{itemize}
    \item \textbf{Enhancement on reliability}:
\end{itemize}
% \textit{Enhancement on reliability.}

In this case study, we explore how the knowledge augmentation phase improves the compatibility between the LLM and a selected model. Without knowledge augmentation, LLMs might generate variables or implement data retrieval logic independently, leading to redundancy and inefficiency. However, with knowledge augmentation, the LLMs are better informed about the existing components and functions within the selected model, enabling them to utilize the model's APIs(in this case study, the model represents Revit) effectively.

As shown in Fig. \ref{RQ3code1}, the code above is generated without external information, while the code below incorporates entity information extracted during the ED phase. The code generated by naive CoT prompting attempts to retrieve model elements by assuming method input parameters or arbitrarily defining variables or methods to represent the required parameters (e.g., ''assuming the user has implemented a variable num\_detector to represent the number of detectors in a room'' or ''assuming the existence of a function CheckDetectorPerRoom() to return the number of detectors''). This approach can sometimes result in compilation errors and a lower pass rate. In contrast, the knowledge-augmented code generation leverages existing Revit APIs and directly generates logic to retrieve model elements, resulting in more efficient and consistent code. This case study demonstrates that knowledge augmentation significantly enhances the compatibility between LLMs and the selected model, ensuring that the LLMs make optimal use of available resources.
% \begin{itemize}
%     \item \textbf{Improvement on the control of granularity in function generation}:
% \end{itemize}
\begin{itemize}
    \item \textbf{Improvement on the control of granularity}:
\end{itemize}
% \textit{Improvement on the control of granularity.}

This case study investigates how knowledge augmentation influences the granularity of functions generated by the LLMs. Granularity control is crucial for maintaining a balance between high-level abstractions and detailed implementations in code generation. Knowledge augmentation equips the LLMs with detailed information about the required level of abstraction, enabling them to generate functions with appropriate granularity.

\begin{figure}[htpb]
\centering
\includegraphics[width=1.0\linewidth]{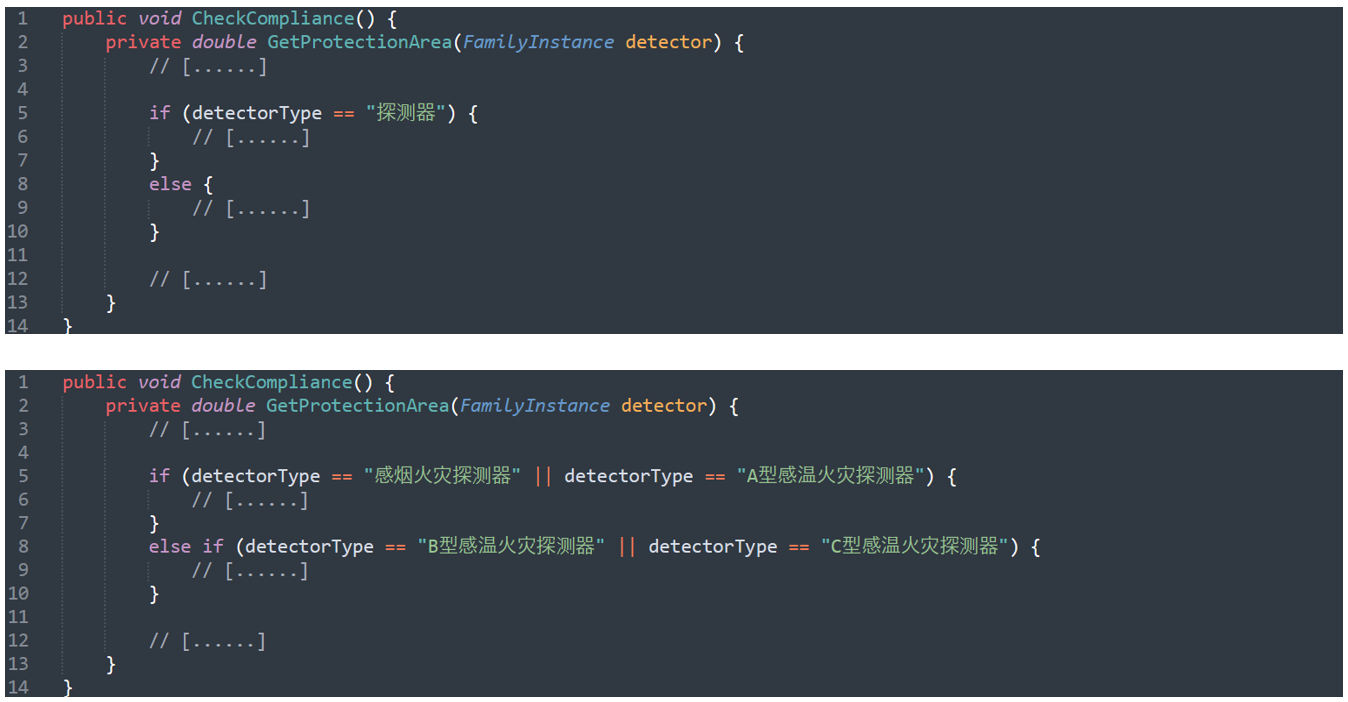}
\caption{An example of granularity in function generation. The code above is generated without knowledge augmentation and the code below is generated with knowledge augmentation}
\label{RQ3code2}
\end{figure}
As illustrated in Fig. \ref{RQ3code2}, the code above is generated without knowledge augmentation and the code below is generated with knowledge augmentation. For instance, for rule entry 6.2.2 in GB50116-2013, which involves different types of detectors (e.g., smoke fire detectors, A and B type temperature fire detectors, etc.), the code generated without the additional knowledge extracted from the IE stage of ARCEAK produces only a simple conditional check, ignoring the logic required to handle various detector types. In contrast, the code generated with knowledge augmentation includes more detailed and accurate handling logic. For complex rule entries like rule entry 6.2.2—which involves more than ten variables as decision criteria and exceeds 1,000 words in total length—LLMs tend to generate only abstract portions of the logic if the content is not explicitly emphasized in the prompts. However, when entity and event information extracted during the IE stage is included, LLMs are constrained to generate code that aligns with the extracted knowledge.  The code generated without The knowledge-augmented LLM generates functions with better-controlled granularity, resulting in clearer, more maintainable code. This case study highlights the role of knowledge augmentation in guiding the LLMs to produce functions that align with project guidelines, thereby improving the overall quality of the generated code.

\begin{tcolorbox}[colback=gray!10,
                  boxrule=.2mm,
                  arc=2mm,
                  auto outer arc,
                  before upper={\parindent15pt\noindent},
                  ]
    \textbf{Answer to RQ3:} By integrating extracted information, ARCEAK enhances the LLM's ability to generate accurate and efficient code. These case studies illustrate that knowledge augmentation not only improves compatibility with existing models but also ensures better control over the granularity of functions. Consequently, the overall quality of the generated code is significantly improved, demonstrating the value of the information extraction phase in the ARCEAK framework.
\end{tcolorbox}

% \subsection{Evaluation Metrics}
\section{Discussion}
\label{sec: discussion}
In this section, we summarize the correct and incorrect cases encountered during the experiments and evaluation process, and analyze the potential causes behind them. We then discuss several threats that may impact the effectiveness of our work.
\subsection{Case Study: Correct and Incorrect Results}
We analyze several cases demonstrating ARCEAK's ability to guide LLMs in generating compliance checking code that better aligns with practical needs. However, there are still instances where our approach does not fully prevent the generation of erroneous results. Below, we provide a brief summary of these situations for further clarification.
\subsubsection{\textbf{Generating API or parameter}}
As shown in Fig.\ref{RQ3code1}, the code below generates the correct API for retrieving Room information: "FilteredElementCollector(doc).OfCategory(BuiltInCategory.OST\_Rooms)". Using the Room API as an example, the API consists of four key components: two element retrieval APIs ("collector" and "filter") and two required parameters ("doc" and "class" enumeration). A correctly generated code for retrieving elements must ensure the accuracy of all four components. First, for the "doc" parameter and its retrieval, GPT-3.5 occasionally uses annotations or custom variables (e.g., "doc = \_doc" or "Room room = getRoom() //implement your logic here to get room"), whereas GPT-4 is more likely to generate the correct "doc" parameter and element retrieval API. Next, with respect to the "class" enumeration, LLMs sometimes generate incorrect class enumeration values, particularly when dealing with complex entities. In such cases, LLMs might create custom enumeration values. Notably, GPT-4 demonstrates greater consistency and accuracy in generating the correct "class" enumeration values compared to GPT-3.5.
\subsubsection{\textbf{Generating verification code for complex rule entry}}
For verification code generation, positive pass rate examples are primarily concentrated on relatively simple rule entries. However, for rule entries involving complex operations (e.g., calculating the distance from a detector to the centerline of a wall), the pass rate begins to decline. GPT-3.5 often attempts to generate methods that are not implemented (without the ⟨unimplemented⟩ tag) to assume data, while GPT-4 tries to generate correct code but may sometimes miss special cases (e.g., transforming walls to avoid calculating the distance to the side of the wall, or excluding the floor where the detector is located when checking for obstructions within a certain range around the detector). For rule entries with complex logic (e.g., rule entry 6.2.2 in GB50116-2013, which contains over 1,000 words and includes multiple nested value intervals), even GPT-4 may overlook part of the decision logic.
\subsection{Threats To Validity}
In this section, We identify two main threats to the validity of our study:
\subsubsection{Limited selection of models.} In this paper, we selected two LLMs for our experiments. However, it is important to acknowledge that other LLMs are available, including general models like Llama3~\cite{Llama3} and specialized models like CodeGen~\cite{DBLP:conf/iclr/NijkampPHTWZSX23}. In future work, we plan to conduct experiments with a broader range of LLMs to more comprehensively explore the applicability of our framework.
\subsubsection{Limited dataset.} In this paper, we used twenty room-level BIM models to verify the correctness of the code generated by ARCEAK, specifically testing the generation of C\# code from text. It remains uncertain whether our experimental results and findings can be generalized to other languages (e.g., generating Python code for checking in Dynamo). In the future, we plan to build a larger building model dataset and make it publicly available to further support ARC research.

\section{Related Work}
Automated Rule Checking (ARC) is a technology-driven approach that automates the compliance verification process by converting building rules into computer-recognizable formats, such as decision tables~\cite{fenves1969decision} or query code~\cite{peng2023automated}. As engineering projects become increasingly complex, manual compliance checking has grown both tedious and costly, while also raising the risk of human errors. In response, ARC offers a solution that can significantly reduce both time and expenses, all while improving the quality and accuracy of reviews. The rise of building modeling technologies has further bolstered this automation by making data more machine-readable, facilitating smoother integration into the compliance checking process. The ARC process consists of three key stages: 1) rule interpretation, which converts natural language rules into machine-readable formats, 2) building model preparation, which organizes the necessary information for rule checking, and 3) rule execution, where the prepared model is checked against the machine-readable rules~\cite{zheng2022knowledge}. Of these stages, rule interpretation and rule execution are particularly crucial and complex, warranting further research~\cite{ismail2017review}.

Current research on ARC for conditional rules still involves considerable manual effort, such as entity labeling or sentence reconstruction \cite{zhong2018ontology, beach2020towards}. Fang et al.~\cite{fang2020knowledge} proposes a knowledge graph that fuses computer vision with ontologies to dynamically recognize construction hazards while adhering to evolving safety standards. Zhou et al.~\cite{zhou2015research} proposes a smart method for diagnosing wind turbine faults using ontology-based FMECA knowledge and a JESS rule engine to speed up maintenance decision-making. Zhou et al.~\cite{zhou2022arccfg} utilized the pretrained language model BERT for automated semantic annotation to capture the semantic information of sentences and then generated code from labeled sentences. Zhen et al.~\cite{zheng2022knowledge} established an ontology to represent domain knowledge and then generate SPARQL-based queries based on a pattern matching algorithm. Even though ready-
to-use NLP tools for knowledge extraction exist, there is still
room for refinement in structured text processing tools to more
closely cater to the unique characteristics of specific fields.
% Despite these advancements, their approaches still require labeled data to achieve optimal performance and have limitations regarding the scope of building rules they can process.

In addition to the challenge of standardizing architectural design rules, another critical factor affecting the efficiency of ARC is the availability and accessibility of data. Zhang et al.~\cite{zhang2016extending} proposes a method that enhances automated compliance checking in building designs by merging compliance information into the IFC schema using machine learning and natural language processing. Although the IFC standard format is often favored for ARC~\cite{melzner2013case}, Malsane et al.~\cite{malsane2015development} points out that Building Information Modeling (BIM) models typically lack the necessary detail level for such checks. Given the interconnected nature of BIM design and code checking, expecting BIM modelers to include all necessary review details is impractical. Many commercial ACC systems still require manual input of certain data. Although machine learning has proven to be useful in semantically enriching and reconciling IFC data exchange issues~\cite{koo2019using}, converting the extensive information present in BIM models into a machine learning-friendly format continues to be an overwhelming task~\cite{sacks2019automating}.

Despite significant advancements in ARC within the AEC industry, extracting requirements and compliance information from detailed text documents remains a challenge. Innovations in NLP and knowledge graphs are improving the efficiency and accuracy of these systems, but the full automation of BIM reviews continues to be an ongoing effort. This paper builds upon existing research and, leveraging LLMs, seeks to transform natural language building rules into executable verification code with minimal human intervention. This marks a step towards the intelligent evolution of compliance checking in construction.

\section{Conclusion}
\label{sec: conclusion}

In this work, we focus on the field of Automated Rule Checking (ARC) with the aim of reducing the manual effort required to convert natural language building rules into verification code for selected models. To achieve this, we propose ARCEAK, a novel LLM-based Automated Rule Checking framework Enhanced with Architectural Knowledge. ARCEAK achieves an almost fully automated process for converting natural language building rules into executable verification code. By consulting construction domain experts, we developed a robust construction domain-specific entity and event schema and established appropriate verification and evaluation metrics. The evaluation results of ARCEAK demonstrate outstanding performance in rule information extraction and an acceptable compile pass rate in verification code generation.

In the future, we plan to continue improving our framework. This includes, but is not limited to, expanding our methodology to cover a wider range of building rules and providing LLMs with related API lists during the Rule Checking Code Completion stage to enhance compatibility between the LLMs and selected models, which is crucial for improving the logic pass rate. Additionally, we aim to evaluate our framework on real architectural blueprints.

\bibliographystyle{IEEEtran}
\bibliography{reference.bib}

\end{document}